\documentclass[11pt]{article}
\usepackage{graphicx}
\usepackage{cs12,html}

\begin{document}
 
\title{MUSICOS 1998: Observations of Rotational Modulation and Flares on the RS CVn Binary HR1099 }
 
\author{D. Garc\'{\i}a-Alvarez\altaffilmark{1}, B.H. Foing\altaffilmark{2}, 
D. Montes\altaffilmark{3}, J. Oliveira\altaffilmark{2}\altaffilmark{,}\altaffilmark{4}, J.G. Doyle\altaffilmark{1}, and MUSICOS 98 collaboration}
\altaffiltext{1}{Armagh Observatory}
\altaffiltext{2}{ESA Space Science Department, ESTEC}
\altaffiltext{3}{Departamento de Astrof\'{\i}sica, Universidad Complutense de Madrid}
\altaffiltext{4}{Centro de Astrofisica da Universidade do Porto}


\begin{abstract}

We present simultaneous and continuous observations of H$\alpha$, 
H$\beta$, \ion{Na}{1} D$_{1}$, D$_{2}$, \ion{He}{1} D$_{3}$  and \ion{Ca}{2} H \& K lines
   of the chromospherically active binary HR 1099.  
   We have observed HR 1099 for more than 3 weeks almost continuously and 
monitored two flares.
   An increase in H$\alpha$ and  \ion{Ca}{2} H \& K, H$\beta$ and 
\ion{He}{1} D$_{3}$  and a
   strong filling-in of the \ion{Na}{1} D$_{1}$, D$_{2}$ during the flares are observed. 
We have found that the flares took place at the same phase
   (0.85) of the binary orbit, and both of them seems to occur near the limb. 

\end{abstract}

\section{Introduction}

HR 1099 is a triple system and consists of a close double-lined spectroscopic pair with a spotted and rapidly rotating K1IV primary and a
comparably inactive and slowly rotating G5V secondary in a $2\fd8$ orbit. The terciary is a fainter K3V star $6\arcsec$ away. The system HR 1099 is one of the few RS CVn systems, along with UX Ari, II Peg, and DM UMa, that shows H$\alpha$ 
consistently in emission. Its tidally induced rapid rotation, combined with the deepened 
convection zone of a post main sequence envelope, is
   responsible for the very high chromospheric activity for its spectral 
class. The primary (K1IV) shows very strong and variable CaII H \& K and 
H$\alpha$
   emission that is indicative of high chromospheric activity.
A comparison of
the close binary with evolutionary models suggests that mass transfer from the K primary onto the G secondary may begin within $10^{7}$ years
(Fekel 1983). 

Following previous MUSICOS campaigns on HR 1099 (Foing et al. 1994) the main goals of this campaign were to monitor for flares and to observe
chromospheric lines in order to diagnose the energetics and velocity dynamics. Also photospheric Doppler Imaging and the study of 
chromospheric activity variations were planned.

\section{Observations}

The 
spectroscopic observations have been obtained during the MUSICOS 1998 campaign (Nov-Dec), using both 
Echelle and Long Slit Spectrographs. The sites and instruments involved in the campaign and some of their most
important characteristics are shown in Table~1. The spectra have been extracted using the standard reduction
procedures in the IRAF package. The wavelength calibration was obtained by taking spectra of a Th-Ar lamp. The
spectra have been normalized by a low-order polynomial fit to the observed continuum. Finally, for the spectra
affected by water lines, a telluric correction have been made.

\begin{table*}[]
\begin{center}
\caption{ }
\begin{tabular}{ccccc}
\hline \\
Site & Telescope & Spectrograph & Number & Resolving  \\
     &           &              & Nights & Power      \\
\hline \\
ESO, Chile & 0.9 m & HEROS & 11 & 20\,000 \\
INT, La Palma & 2.5 m & ESA-MUSICOS & 9 & 35\,000  \\
KPNO, USA & 0.9 m & Echelle & 10 & 65\,000 \\
LNA, Brazil & 1.6 m & Coud\'{e} & 6 & 60\,000  \\   
Mt.Stromlo, Australia & 1.9 m & Echelle & 5 & 35\,000 \\
OHP, France & 1.9 m & Elodie & 8 & 43\,000  \\
OHP, France   & 1.5 m & Aurelie & 6 & 22\,000  \\
SAAO, South Africa & 1.9 m & Giraffe & 3 & 36\,500  \\
Xinglong, China & 2.2 m & Echelle & 10 & 35\,000  \\

\hline \\
\hline
\end{tabular}
\end{center}
\end{table*}

\section{Results}

\begin{figure}[t]
\begin{center}
\includegraphics[angle=0, width=10cm]{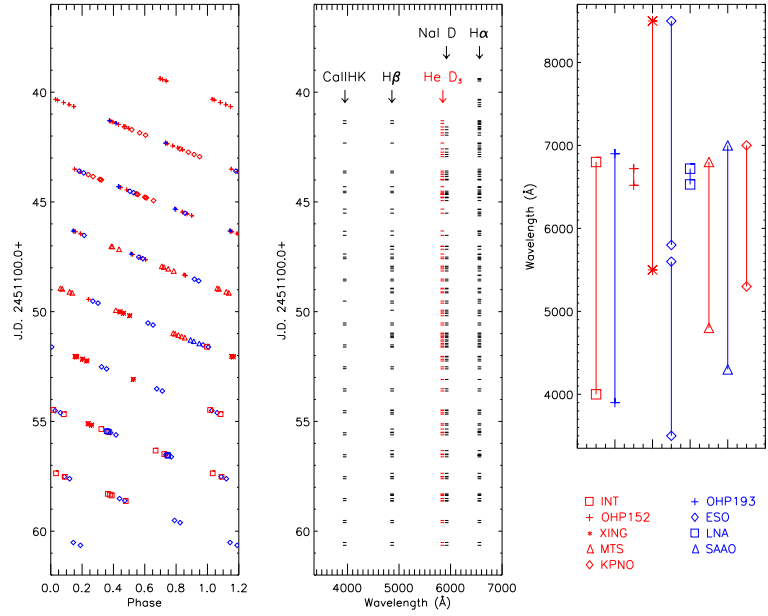}
\end{center}
\hspace{1.75cm}
\includegraphics[angle=0, width=10cm]{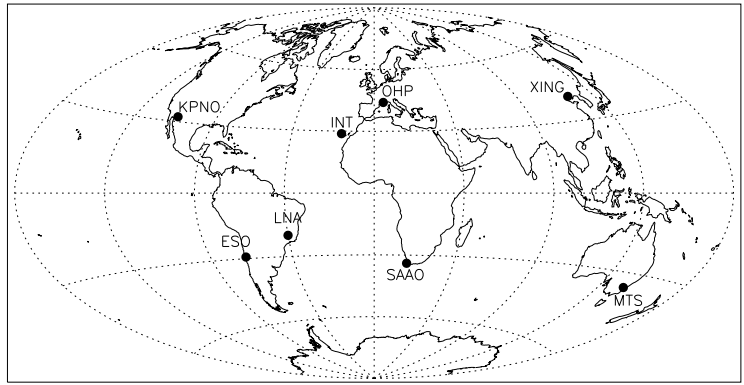}

\caption{Top left: Phase coverage of the spectroscopic
observations versus Julian date. Top middle: Line coverage of the observations of HR 1099 during
the MUSICOS 98 campaign. Top right: Wavelength
coverage of the instruments involved in the MUSICOS 98 campaign. Bottom: The sites involved in the
MUSICOS 98 campaign: INT, OHP, XING (Xinglong), KPNO (Kitt Peak), MTS (Mt. Stromlo), LNA, ESO
and SAAO.}
\label{FigCoverage}

\end{figure}

Phase coverage, line coverage, wavelength coverage and sites involved in the MUSICOS 1998 campaign are shown in Fig~\ref{FigCoverage}. 
From the sites involved in the campaign we can conclude that HR 1099 was observed almost continuously during the campaign, approximately 3
weeks. 

\begin{figure}[t]
\begin{center}
\includegraphics[angle=0, width=10cm]{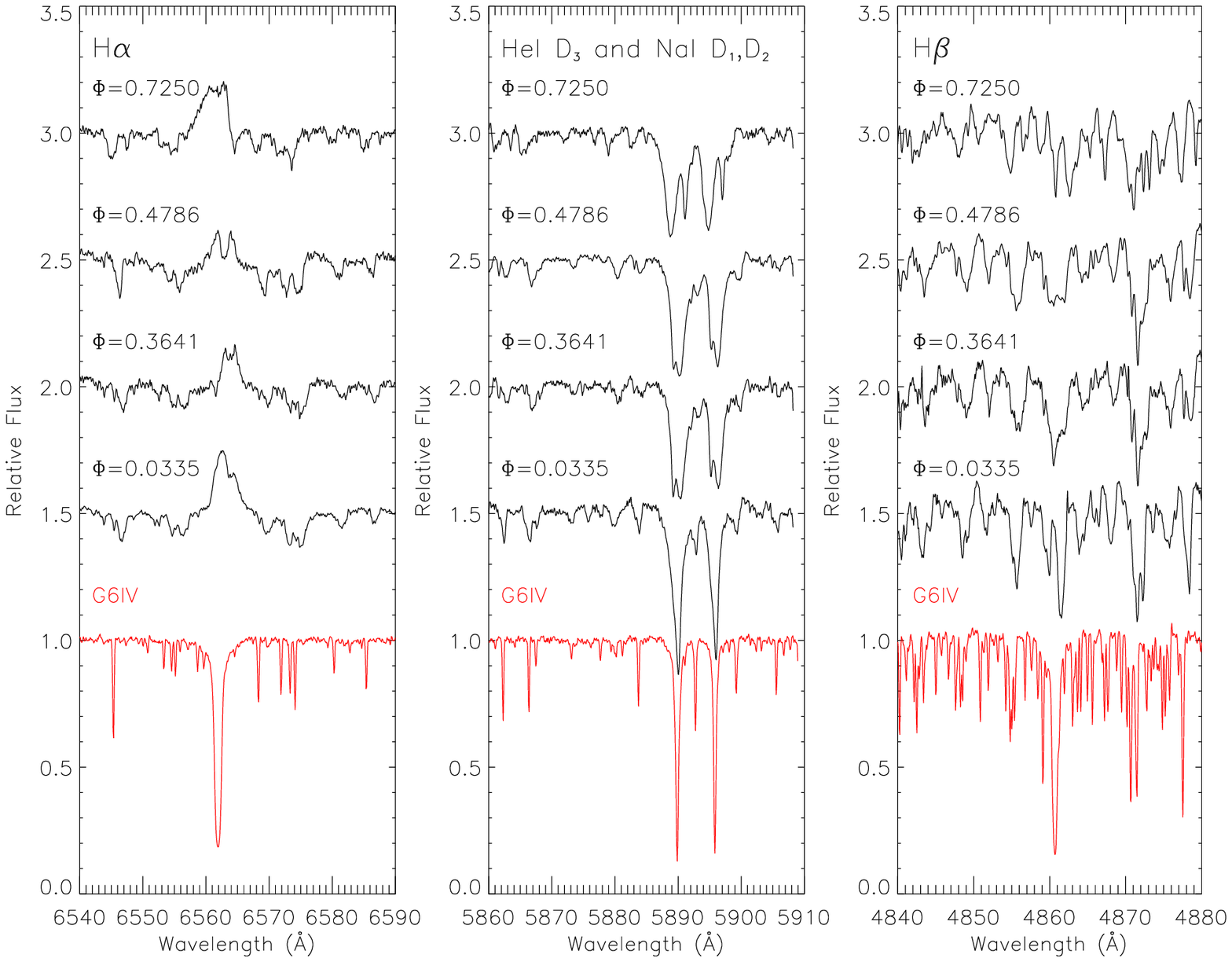}
\end{center}
\hspace{1.75cm}
\includegraphics[angle=0, width=10cm]{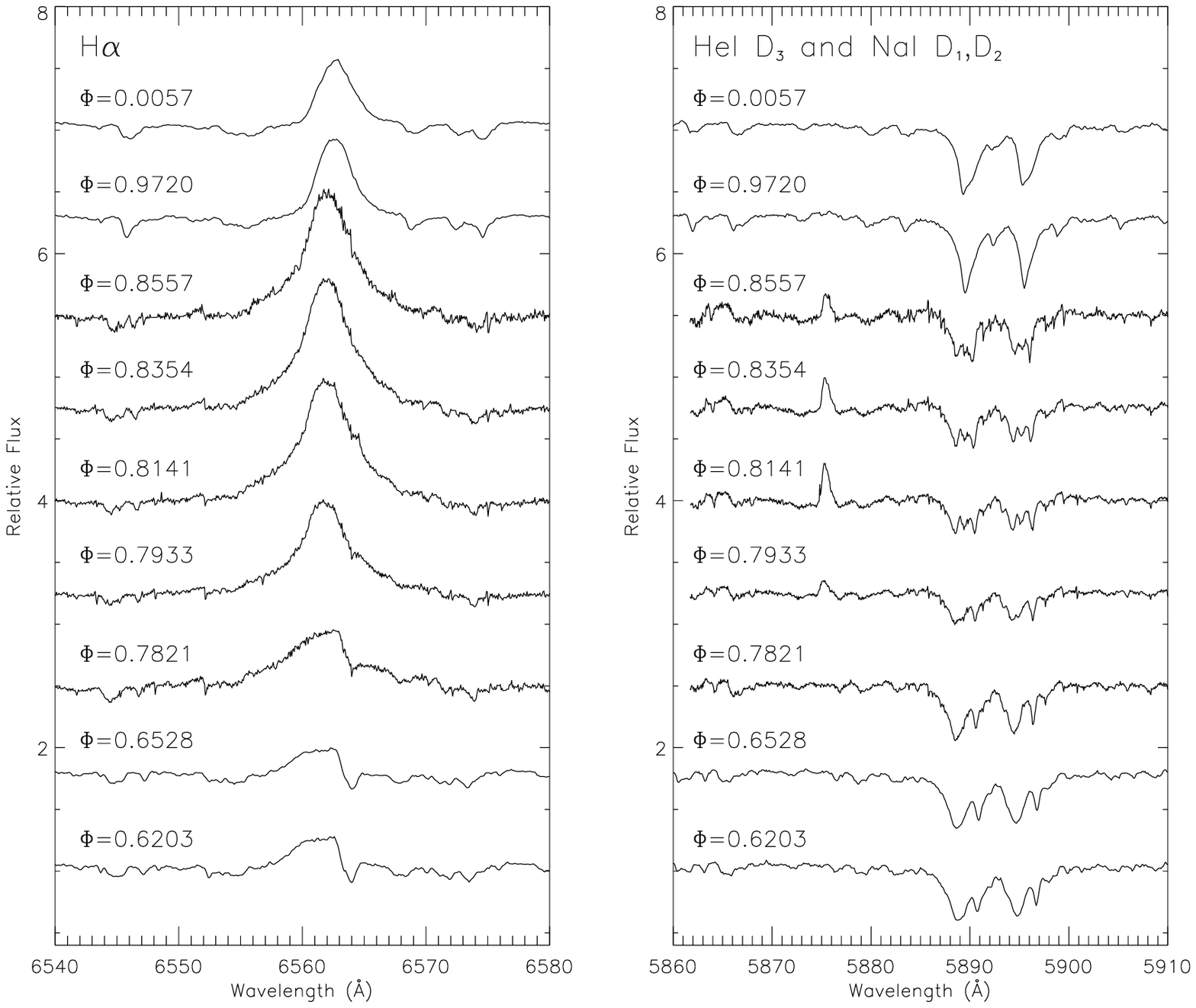}

\caption{Top panels: Phase sequence of optical spectra observed during the MUSICOS 1998 campaign. A
spectral standard type G6IV spectrum is shown for comparison. Bottom panels: The observed spectra for 
H$\alpha$ (left panel) and \ion{He}{1} D$_{3}$ and \ion{Na}{1} D$_{1}$,D$_{2}$ (right panel) of the second monitored flare 
   starting at JD 2451151.066 arranged in order of the orbital phase.  }
\label{FigSample}

\end{figure}

Sample spectral of HR 1099 showing the chromospheric lines, namely, H$\alpha$, H$\beta$, \ion{Na}{1} D$_{1}$, D$_{2}$, 
\ion{He}{1} D$_{3}$, at different phases are presented
in Fig~\ref{FigSample} (top panel). During the campaign two flares were observed, one at JD 2451145.513 (28-11-98) and a second flare at JD
2451151.066 (03-12-98). This second flare shows 
an increase in the H$\alpha$ emission line, \ion{He}{1} D$_{3}$ line turns into emission during
 the flare and it also shows a strong filling-in of the \ion{Na}{1} D$_{1}$, D$_{2}$ line (see bottom Fig~\ref{FigSample}). 

\begin{figure}[t]
\begin{center}
\includegraphics[angle=0, width=10cm]{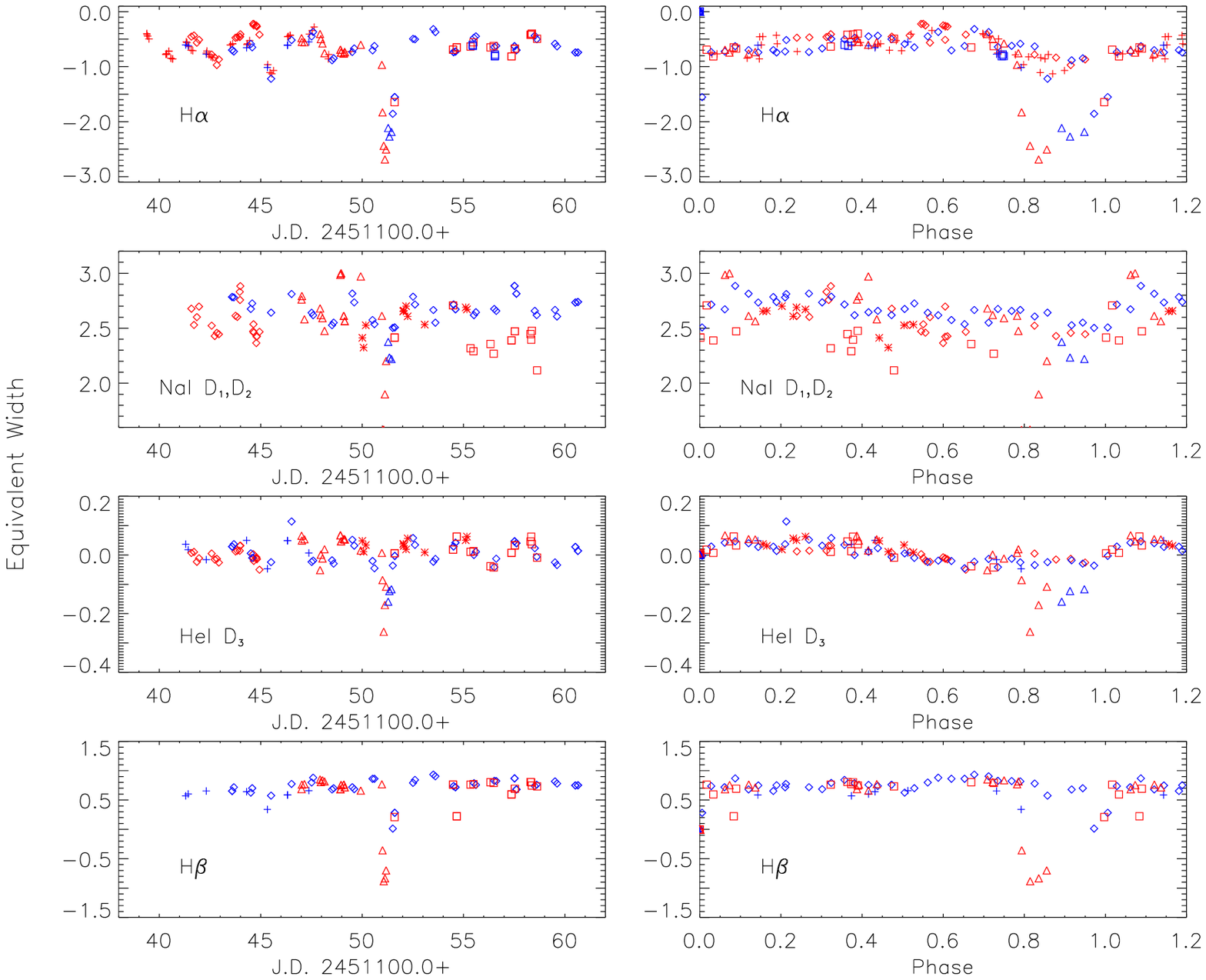}
\end{center}
\hspace{1.75cm}
\includegraphics[angle=0, width=10cm]{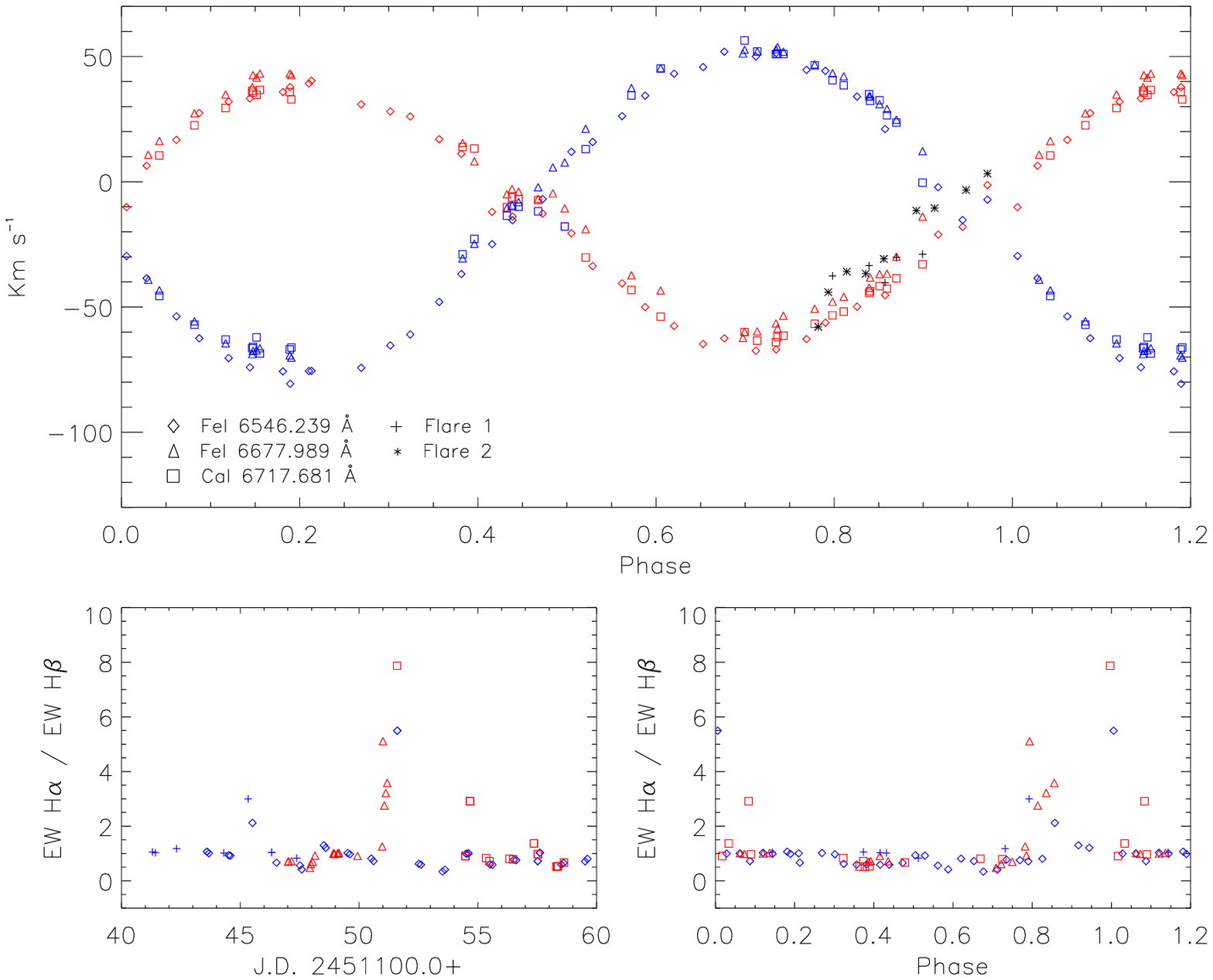}

\caption{Top panels: The equivalent width as a function of Julian date and as a function of phase for H$\alpha$, \ion{Na}{1} D$_{1}$,D$_{2}$
, \ion{He}{1} D$_{3}$ and H$\beta$ lines. Middle panel: The radial velocity curves 
of the binary systems HR 1099, calculated using photospheric lines. Red and blue symbols are used for the primary star and secondary star respectively.  Bottom panels: The 
 $\rm{EW_{H\alpha}/EW_{H\beta}}$ ratio as a function of Julian date and as a
 function of phase. Symbols follow Fig.1. }
\label{FigEquiWidth}

\end{figure}

In the top panel of Fig~\ref{FigEquiWidth} we show the equivalent width as a function of Julian date and as a function of phase for several chromospheric
lines. From these results we observed that, although both flares
shown an increase in H$\alpha$ emission, the second flare produced a bigger increase. We have also observed that the first flare shows
 filling-in of H$\beta$ but for the second flare this line turns into emission. Another remarkable feature is that the \ion{Na}{1} D$_{1}$, 
 D$_{2}$  lines have a strong filling-in during the second
 flare. Note that both flares took place at around the same phase (0.85), but $\sim$6 days apart. Note also that we have detected rotational
 modulation of the \ion{He}{1} D$_{3}$ line that could be attributed to the pumping of the HeI line by coronal X-rays
 from active regions.

In the middle  panel of Fig~\ref{FigEquiWidth} we plot the radial
  velocity curves of the binary system HR 1099, calculated using photospheric lines. We have also plotted the radial velocity for the two
  monitored flares, calculated using H$\alpha$. As a result, we notice that during both
 flares, the radial velocity is slightly displaced (15-20 $\mathrm{km\ s^{-1}}$) compared to the center of
 gravity of the primary. This result could be due to the fact that both flares had taken place near the limb. 

The bottom panel of Fig~\ref{FigEquiWidth} shows the 
 $\rm{EW_{H\alpha}/EW_{H\beta}}$ ratio as a function of Julian date and as a
 function of phase. During quiescent we obtained values around 1 for the $\rm{EW_{H\alpha}/EW_{H\beta}}$ ratio, while during the first flare 
 we obtained values slightly bigger than in the quiescent (2-3). During the second flare the ratio even reach values of 8.

\section{Conclusions}

We have observed the binary system HR 1099 continuously for more than seven orbits in the optical
spectral range (at high spectral resolution over a wide wavelength domain) and we have also monitored 
two flares,  one of which lasted more than one day.

Flares spectroscopic variations were measured in H$\alpha$, with H$\beta$ and \ion{He}{1} D$_{3}$ turning into 
emission and a strong filling-in of the NaI Doublet.

We have detected a rotational modulation of the \ion{He}{1} D$_{3}$ line outside flares, that may indicate non
axi-symmetry
in the distribution of coronae active regions.

We shall compare these results with Doppler Imaging based on the photospheric lines, to
study the connection between spots, chromospheric emission and flares.

\acknowledgments

      We wish to thanks those that have contributed to the MUSICOS 1998 campaign. Research at Armagh Observatory is grant-aided by the 
      Department of Culture, Arts and Leisure for Northern Ireland. DGA wished to thank the Space Science Department at ESTEC the finacial
      support. DM is supported by the Spanish DGESIC under grant PB97-0259.


\end{document}